\documentclass[conference]{IEEEtran}

\usepackage{graphicx}
\usepackage{url}
\graphicspath{{./Figures/}} 
\usepackage{subfigure}
\usepackage{amsmath}

\usepackage{algorithm}
\usepackage{algorithmicx} 
\usepackage{algcompatible}

\usepackage{scalerel,stackengine}

\usepackage{url}
\usepackage{breakurl}

\usepackage{hyperref}
\newcommand\fnurl[2]{%
\href{#2}{#1}\footnote{\url{#2}}%
}
\newcommand\blfootnote[1]{%
  \begingroup
  \renewcommand\thefootnote{}\footnote{#1}%
  \addtocounter{footnote}{-1}%
  \endgroup
}

\stackMath
\newcommand\reallywidehat[1]{%
\savestack{\tmpbox}{\stretchto{%
  \scaleto{%
    \scalerel*[\widthof{\ensuremath{#1}}]{\kern-.6pt\bigwedge\kern-.6pt}%
    {\rule[-\textheight/2]{1ex}{\textheight}}
  }{\textheight}%
}{0.5ex}}%
\stackon[1pt]{#1}{\tmpbox}%
}

\begin{document}
%
\title{Deep-Gap: A deep learning framework for forecasting crowdsourcing supply-demand gap based on imaging time series and residual learning}

\author{\IEEEauthorblockN{Ahmed Ben Said, Abdelkarim Erradi}
\IEEEauthorblockA{Department of Computer Science and Engineering, College of Engineering\\
Qatar University, Doha, Qatar\\
Email: \{abensaid, erradi\}@qu.edu.qa}}

\maketitle

\begin{abstract}
Mobile crowdsourcing has become easier thanks to the widespread of smartphones capable of seamlessly collecting and pushing the desired data to cloud services.
However, the success of mobile crowdsourcing relies on balancing the supply and demand by first accurately forecasting spatially and temporally the supply-demand gap, and then providing efficient incentives to encourage participant movements to maintain the desired balance.
In this paper, we propose Deep-Gap, a deep learning approach based on residual learning to predict the gap between mobile crowdsourced service supply and demand at a given time and space. The prediction can drive the incentive model to achieve a geographically balanced service coverage in order to avoid the case where some areas are over-supplied while other areas are under-supplied. This allows anticipating the supply-demand gap and redirecting crowdsourced service providers towards target areas. Deep-Gap relies on historical supply-demand time series data as well as available external data such as weather conditions and day type (e.g., weekday, weekend, holiday). First, we roll and encode the time series of supply-demand as images using the Gramian Angular Summation Field (GASF), Gramian Angular Difference Field (GADF) and the Recurrence Plot (REC). These images are then used to train deep Convolutional Neural Networks (CNN) to extract the low and high-level features and forecast the crowdsourced services gap.\newline
We conduct comprehensive comparative study by establishing two supply-demand gap forecasting scenarios: with and without external data. Compared to state-of-art approaches, Deep-Gap achieves the lowest forecasting errors in both scenarios.
\end{abstract}

\begin{IEEEkeywords}
Crowdsourced service, Supply-demand gap, Time series, Gramian Angular Field, Recurrence Plot, Residual learning 
\end{IEEEkeywords}

\section{Introduction}
The ubiquity of mobile devices and their widespread usage have lead to the emergence of new ways for reaching others and acquiring data by relying on crowd participation. For example, sensors embedded on smartphones allows the general public to report traffic conditions and weather forecasting via a simple mobile application, thus the term mobile crowdsourcing \cite{azadeh_book}. The latter has been extensively studied and several applications and platforms have been implemented and tested particularly in context of smart cities. Chen \textit{et al}. \cite{parking} developed a mobile application to track available parking spots in cities by relying on crowd participation. The authors draw key guidelines: the coordination between participants is very important. Also, it is very critical to maintain high participation rate for successful crowdsourcing system. Dong \textit{et al}. \cite{fuel} proposed a mobile application that allows tracking fuel price across the city using crowdsourcing approach. Whenever a participant is at the vicinity of a petrol station, the mobile camera is triggered. Then, fuel price is extracted using an image processing algorithm. MySanJose App. \footnote{\url{https://www.sanjoseca.gov/mysanjose}} allows San Jose residents to report any street issue such as light problem or pothole. Similarly, NYC311 mobile App. \footnote{\url{https://www1.nyc.gov/nyc-resources/service/5460/nyc311-mobile-app}} offers New York city residents a platform to report problems such as snowy streets and potholes. \newline
Thanks to the ease of access, low cost and wide availability of cloud services, the interoperability between crowdsensors and the cloud has elicited the emergence of a new paradigm called crowdsourced service \cite{ahmed_icsoc,ahmed_mone, access-2019-bensaid}. In this context, the sensor (e.g. smartphone) represents a crowdsourced service with functional and non-functional attributes. The functional attributes model the space and/or time properties of the service such as location and time availability. The non-functional attributes refer to the type of the offered service  and its associated quality of service (QoS). Neiat \textit{et al}. \cite{azadeh_icws} established a scenario from public transport service in which sensors embedded in the public transit vehicle, such as bus or tram, are abstracted on the cloud as sensor cloud service. The sensor cloud service is characterized by its spatiotemporal aspect as it is available at a given space and for a given time. Authors proposed a spatiotemporal A$^{*}$-based algorithm derived from the well known A$^{*}$ shortest path finding algorithm \cite{a_star} to compose between sensors cloud services. In \cite{azadeh_book}, authors proposed an abstraction of WiFi coverage as a crowdsourced service. The Internet access is offered by a participant through his smartphone. In this context, the smartphone is abstracted on the cloud as a service with spatiotemporal aspect. The space is confined by the WiFi signal range while the time is limited by the start and end time of WiFi sharing. \newline
As emphasized by multiple studies such as \cite{parking} and \cite{azadeh_toit}, the level of participation in any crowd platform is a determinant factor for its success. Thus, operating an effective incentive model is of paramount importance in order to drive and increase crowd participation. Mobility is an intrinsic property of a crowdsourced service since it relies on moving crowd. This property offers great opportunity for the service operators to serve multiple locations at multiple times. Following the WiFi coverage example, an incentive model may consist of participation credit that can be used in future whenever the participant is looking for WiFi access through crowdsourced service. Neiat \textit{et al}. \cite{azadeh_toit} addressed the problem of supply-demand of crowdsourced service. Authors  proposed a spatiotemporal incentive approach with the goal of achieving geographically balanced service coverage  in terms of supply and demand. The incentive model considers multiple parameters such as location entropy, time of the day and spatiotemporal density. The incentive model enables redistributing crowdsourced service providers in a way to avoid areas with under-supply while others are over-supplied. As large number of participants and  requesters become involved in the crowdsourced service platform, several issues arise. For example, crowdsourced service providers tend to be concentrated in areas where there is potentially higher demand such as city center or at attraction areas. Therefore, it is very important for the crowd platform manager to plan the supply-demand at each location in advance in order to maximize user satisfaction and thus the quality of experience (QoE) and profit. If one would dispose of a \textit{supply-demand gap prediction} mechanism that accurately determines the supply-demand gap in future time slot at a given location, crowdsourced service platform manager will  provide attractive incentives to service providers to move to the under-supplied areas.\newline
In this paper, we exploit recent advances in data analytics and deep learning to predict the supply-demand gap of a crowdsourced service in a geographic area. 
Our approach uses historical service supply-demand gap to predict the future value using time series analysis. The novelty consists of rolling and encoding the time series as three different types of images. Then, a deep Convolutional Neural Network (CNN) is trained on each image to extract features. Furthermore, we augment the deep CNN network with an additional network that learns from external data such as weather condition (temperature, humidity ...), day type (weekend, weekday, holiday). The trained deep neural network is used to forecast the future supply-demand gap at given location and time.\newline
Our contributions can be summarized as follows:
\begin{itemize}
    \item We propose a scenario using crowdsourced WiFi coverage service in which we aim at achieving geographically balanced coverage areas by anticipating the supply-demand gap using CNN-based forecasting model.
    \item We formulate the prediction approach of supply-demand gap as time series forecasting. We propose to encode the time series as three different types of image. A three pathways deep CNN network is trained on the images. This network is augmented with an additional deep network trained on external data for better prediction performance.
    \item We present a comparative study using real-world datasets to validate the performance of Deep-Gap. We also provide comparison with state-of-art prediction approaches.
\end{itemize}

\section{Motivation Scenario}
\begin{figure}[t!]
	\centering
	\includegraphics[scale=.3]{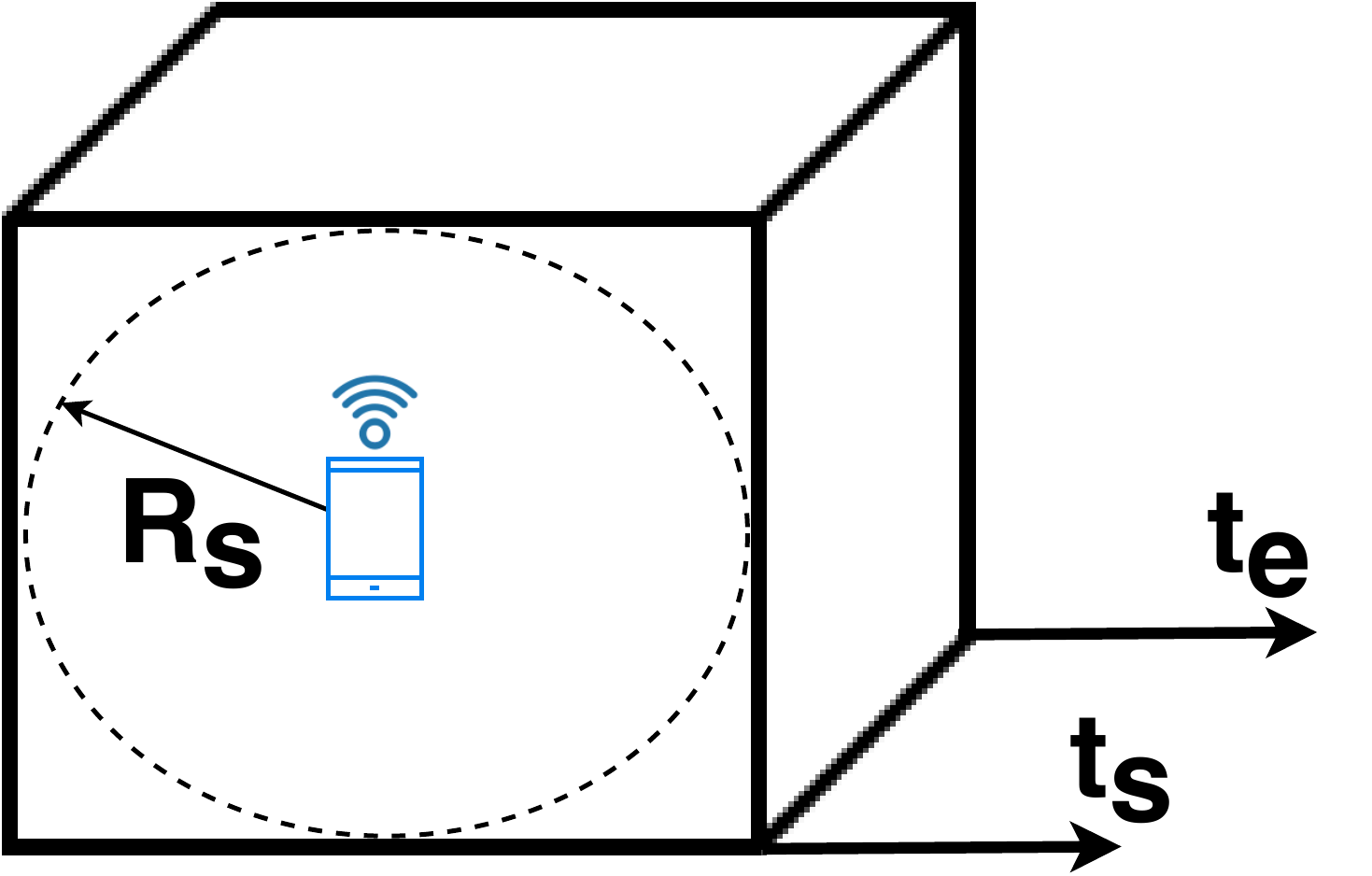}
	\caption{Crowdsourced hotspot service: the coverage area is limited in space by the WiFi signal extent and in time by the start and end time of WiFi sharing.}
	\label{service}
\end{figure}

In this section, we derive a scenario in which we consider crowdsourced WiFi sharing as a service. \newline
WiFi sharing service is represented by a smartphone providing internet access via WiFi tethering. Such feature is available and easily accessible with Android OS. The smartphone is abstracted on the cloud as a service, as illustrated in Fig. \ref{service} with functional and non-functional attributes. 
A crowdourced WiFi hotspot service S has functional and non-functional attributes modelled as a tuple \cite{azadeh_book,access_paper} $<ID, Sensors, space-time,\mathcal{F}, \mathcal{Q}>$:
\begin{itemize}
        \item ID: unique identifier.
        \item Sensors: set of sensors. We suppose that the service consists of one sensor at location $loc$ and sensing area of radius $R_s$.
        \item Space-time: spatio-temporal domain of S. The space is described by a square representing the minimum bounding box of the coverage area as illustrated in Fig. \ref{service}. The time is a tuple ($t_s, t_e$) where $t_s$ and $t_e$ are the start and end time at the current sensing area.
        \item Set of functionalities $\mathcal{F}$: describes the offered service e.g. providing WiFi access.
        \item Set of QoS properties $\mathcal{Q}$ = $(q_1, q_2, ..., q_n)$: where $q_i$ is a QoS property such as the available bandwidth and the WiFi coverage.
    \end{itemize}
    
\begin{figure}[h!]
	\centering
	\includegraphics[scale=.6]{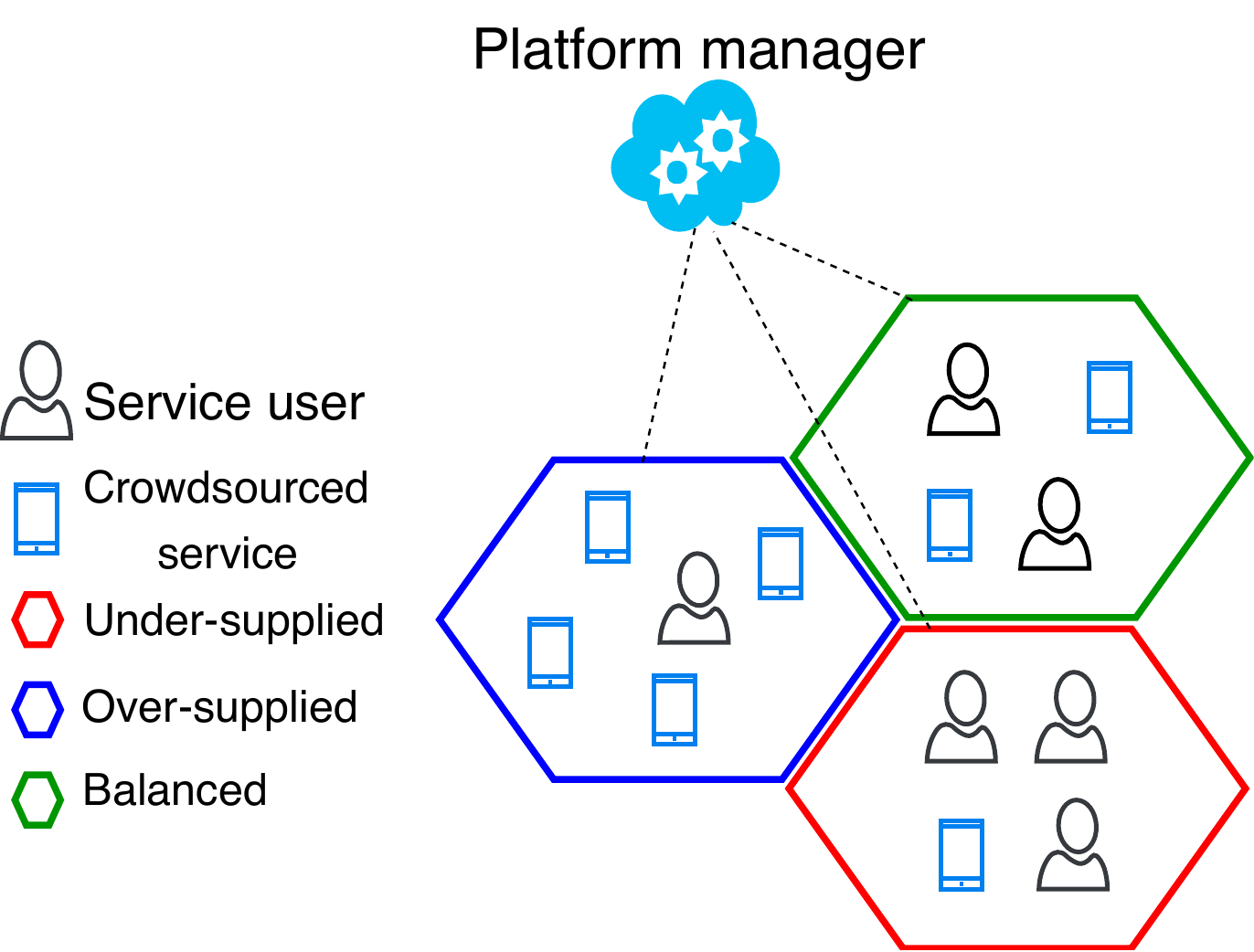}
	\caption{Crowdsourced service platform: The manager is responsible for incentivizing crowdsourced services to move from over-supplied regions to under-supplied regions in order to achieve a balance in terms of supply-demand.}
	\label{scenario}
\end{figure}

Let us consider the scenario illustrated in Fig. \ref{scenario}. The region of study is divided into different geographic areas. At each area, we have a set of crowdsourced services available to share WiFi along with service users looking for WiFi access. In this scenario, we may witness, for a given area, the case where the number of providers is higher than the number of users. This area is said to be \textit{over-supplied}. On the other hand, if the number of users is higher than the number of service providers, the area is said to be \textit{under-supplied}. Otherwise, the area is \textit{balanced}. The crowdourcing success relies on incentivizing crowdsourced services to participate as well as move to under-supplied areas. In order to maximize the Quality of Experience (QoE), it is paramount to ensure that users are served as fast as possible by  nearby crowdsourced services. Furthermore, it is desirable to provide users with failover services to switch to, in case a service is providing unsatisfactory QoS or no longer available. Hence, a solution is needed to anticipate the supply-demand gap at each monitored area and drive the incentive model to maintain a balanced service provisioning and maximize the QoE offered by the crowdsourcing platform.


In this paper, we advocate applying recently proposed  machine learning and data analytics techniques to predict \textit{supply-demand gap}.\newline
We detail in the next section, our formulation of the gap prediction approach using an end-to-end time series forecasting framework.

\section{Supply-demand gap prediction}
We define the supply-demand gap of crowdsourced service at a given area and at a given time slot as the difference between the number of users requesting WiFi access from crowdsourced service and the number of available crowdsourced services. \newline
In this section, we detail the formulation of crowdsourced service supply-demand gap prediction as time series forecasting using the Gramian Angular Summation Field (GASF), Gramian Angular Difference Field (GADF) and Recurrence Plot (REC).

\subsection{Deep neural network architecture}
\begin{figure*}[t!]
	\centering
	\includegraphics[scale=.65]{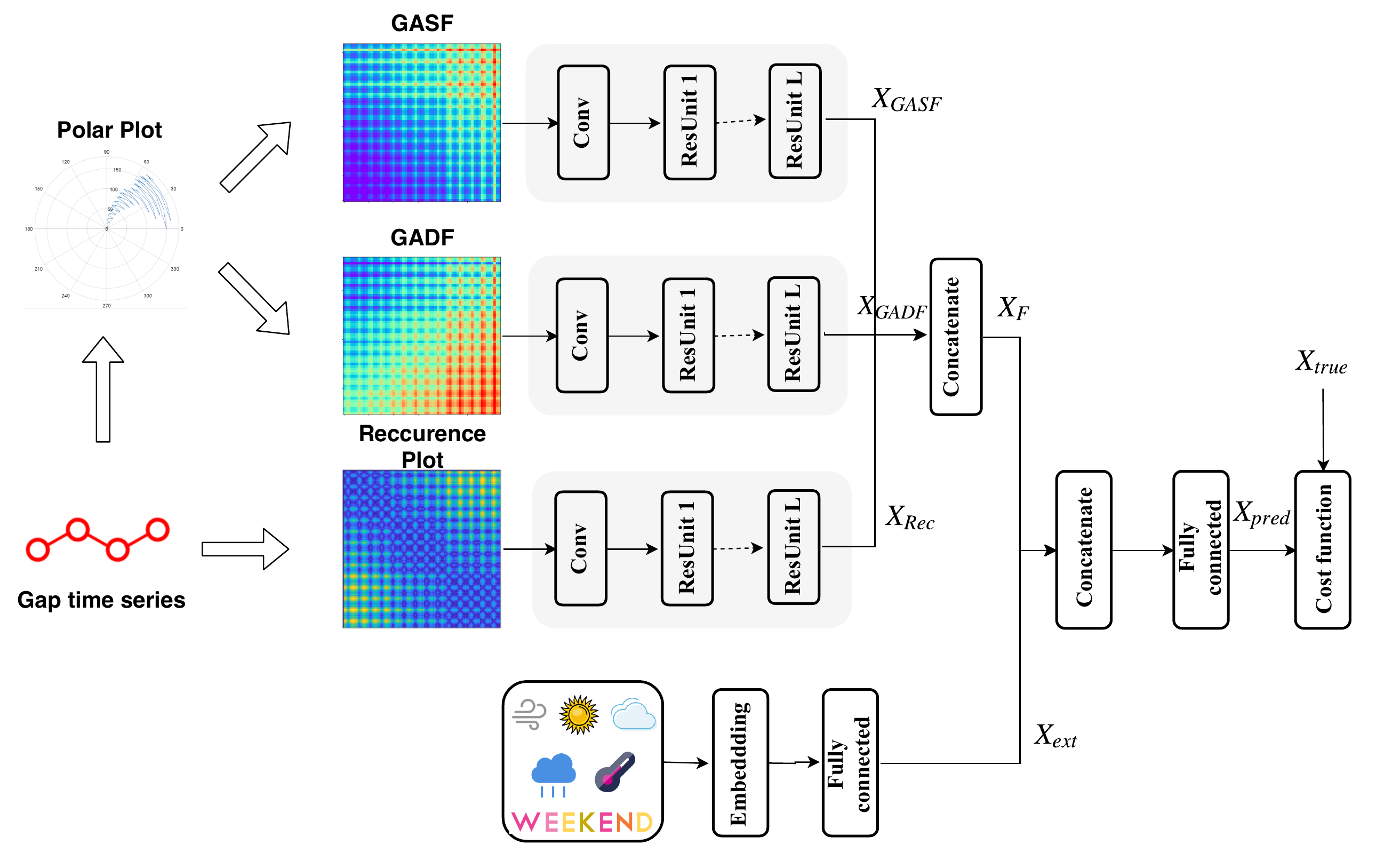}
	\caption{Deep-Gap: Supply-demand gap prediction network: Time series is transformed into polar coordinates and encoded as two types of images: Gramian Angular Summation Field (GASF) and Gramian Angular Difference Field (GADF). The Recurrence Plot is obtained from the original time series. Three pathways deep CNNs are trained on the images and merged with the output of an additional network trained on the external weather and day type data. The overall network is trained to minimize the error between the actual supply-demand gap and the predicted value.}
	\label{network}
\end{figure*}

Fig. \ref{network} depicts Deep-Gap, the proposed supply-demand gap forecasting network. The objective is as follows: \textit{given the time series of supply-demand gap till time slot $T$, predict the supply-demand gap at $T+1$}. A time series of supply-demand gap is defined as $Gap = \{gap_1, gap_2,...,gap_{T}\}$ where $gap_i$ is the supply-demand gap at time slot $i$. Deep-Gap consists of two types of deep neural network. The first one is a  network of three pathways. Each pathway consists of a convolution layer followed by $L$ blocks of Residual Unit \textit{ResUnit}. The first pathway is trained on the GASF image, the second one on the GADF image while the third one is trained on the REC image. The outputs of the pathways i.e. $X_{GASF}$, $X_{GADF}$ and $X_{Rec}$ are then concatenated to obtain $X_F$. The second type of network is dedicated to extract features and learn from external data such as weather condition and temperature. The outputs of both networks, i.e. $X_F$ and $X_{ext}$, are then concatenated and fed into series of fully connected layers. The overall network is trained to minimize a cost function that models the error between the network output $X_{pred}$ and the actual supply-demand gap $X_{actual}$.

\subsection{Imaging time series}\label{ts_image}
\begin{figure}[h!]
	\centering
	\includegraphics[scale=.5]{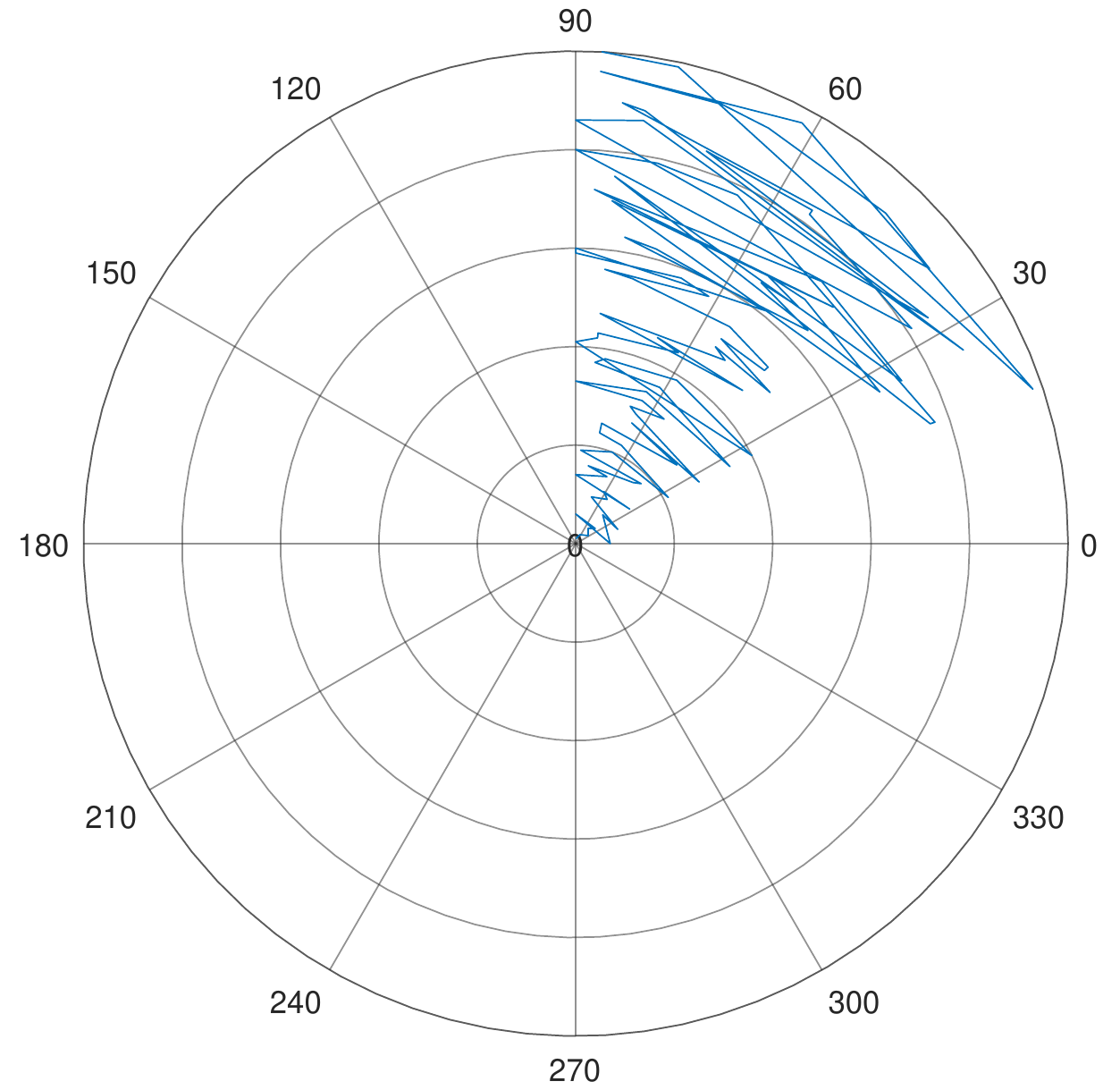}
	\caption{Polar plot of supply-demand gap}
	\label{polar}
\end{figure}

\begin{figure}[!h]
 \centering
 \mbox{
  \subfigure[\label{gasf}]{\includegraphics[scale=0.175]{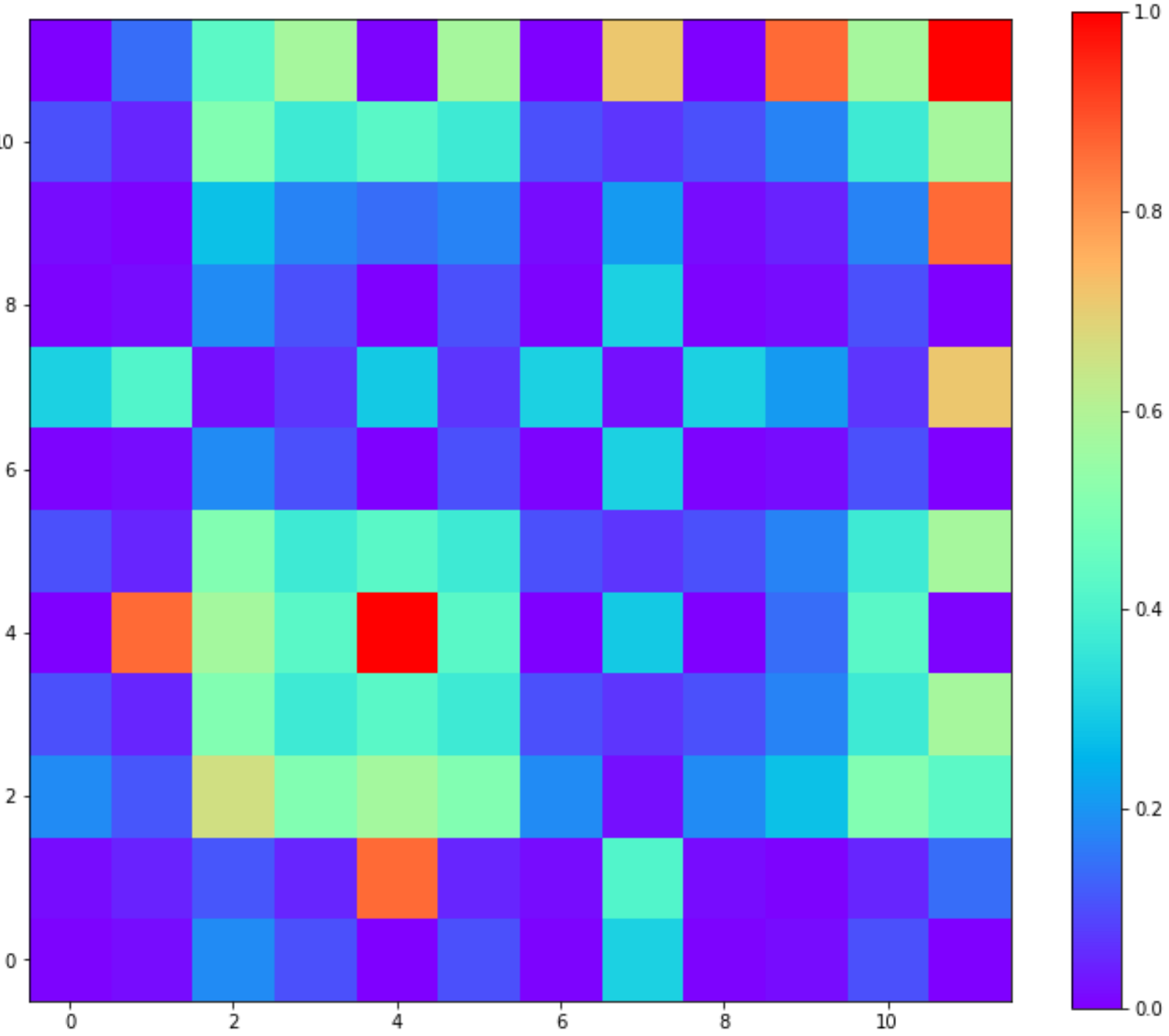}}
  \subfigure[\label{gadf}]{\includegraphics[scale=0.17]{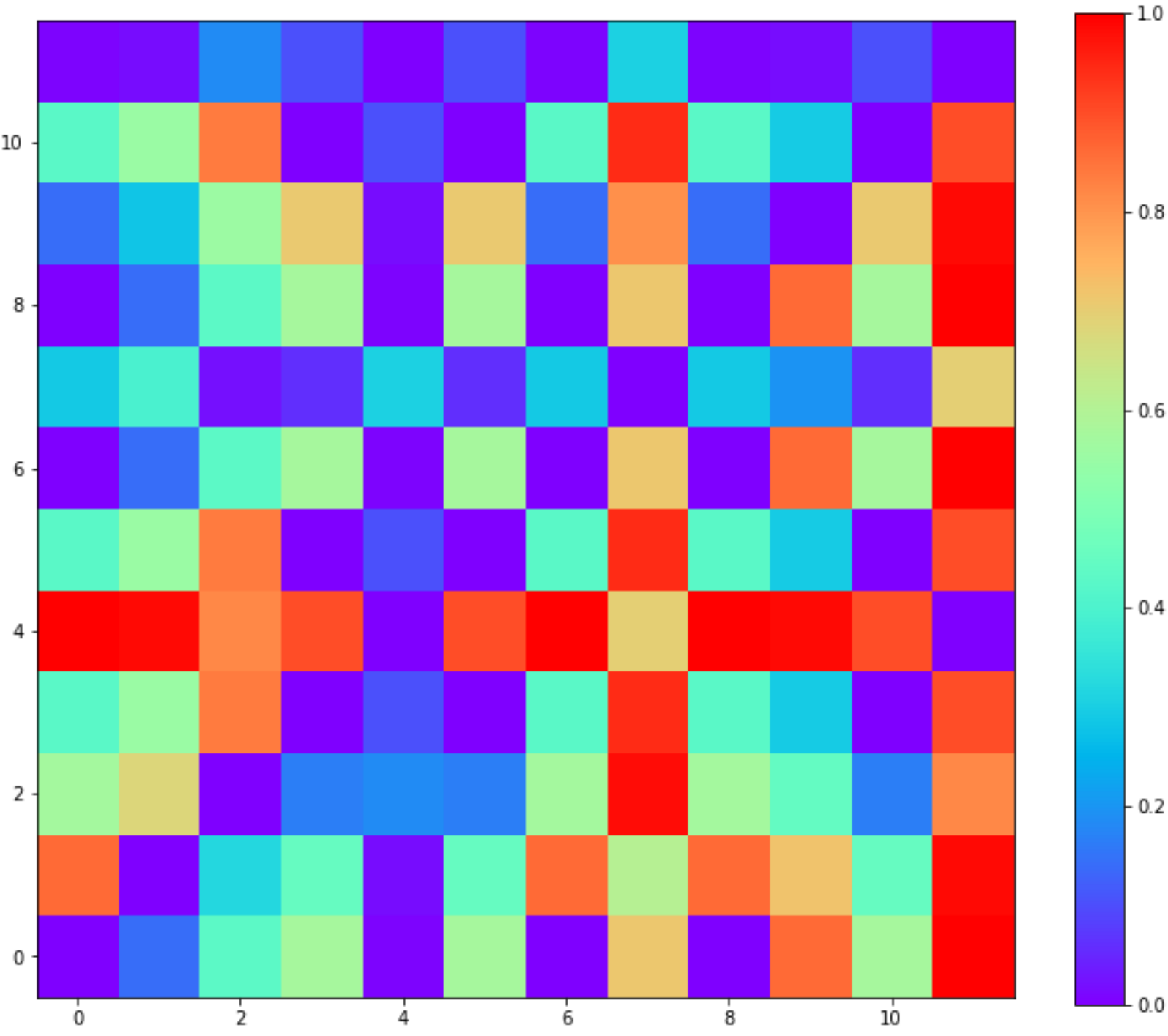}}
  \subfigure[\label{rp}]{\includegraphics[scale=0.175]{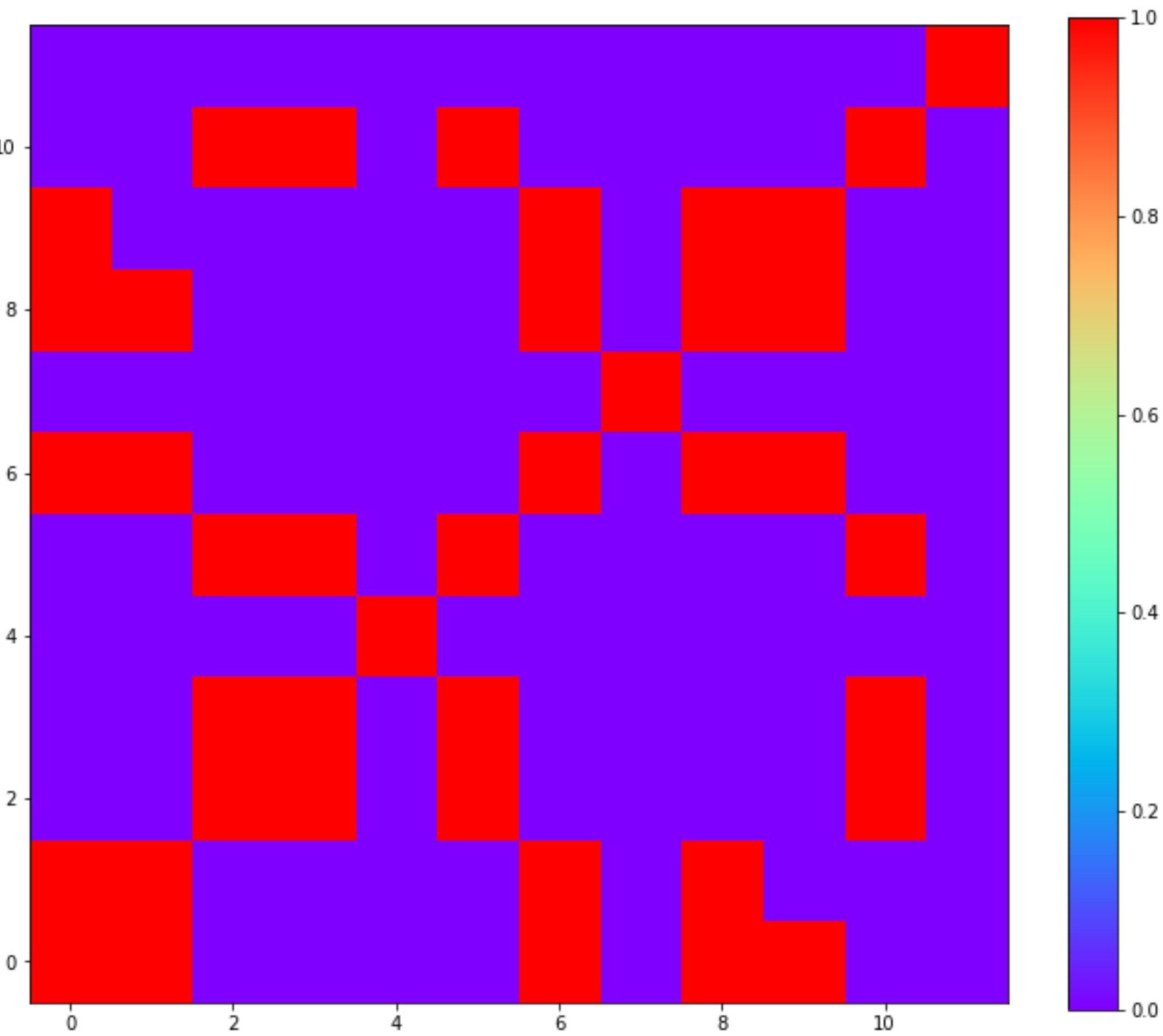}}
 }
 \caption{Imaging time series (a) GASF (b) GADF (c) Recurrence Plot}
 \label{rp_dm}
\end{figure}

In this section, we present our approach to encode time series as images in order to leverage state-of-art deep learning techniques to forecast the $T+1$ gap at a given location.\newline
The idea of transforming time series into other domains has been successfully used in wide range of applications. Indeed, this transformation enables exploiting not only the intra-correlation of the time series but also the inter-correlation. For example, the electrocardiogram (ECG) signal exhibits intra-beat and inter-beat correlations \cite{ecg_corr}. Imaging such biomedical signal has contributed in enhancing the performance in compression applications. In \cite{ecg_corr}, the authors preprocessed ECG signal by segmenting the time series and arranging the resulting segments in 2D. The authors showed that this transformation resulted in better compression performance compared to the 1D counter part. For electroencephalogram (EEG) signal, quantifying long term correlation is challenging. By arranging the EEG time series in 2D form, this correlation is well exploited for multitude of applications including compression and classification of brain waves \cite{eeg_2d}. Srinivasan et al. \cite{eeg_2d_1} proposed a particular arrangement of the EEG time series for compression applications. The transformation consists of first segmenting the signal in equal size blocks. Then the blocks are arranged in a matrix form where odd rows are directly filled while even rows are filled in backward fashion. This zigzag arrangement enables exploiting the correlation of the signal entries for improving EEG compression. In \cite{rpm}, the authors proposed Relative Position Matrix, a novel 2D transformation of time series. First, the dimensionality of the times series is reduced using Piecewise Aggregation Approximation \cite{paa}. Then, the relative position between two time stamps is calculated to form the 2D transformation of the time series. This transformation enables exploiting the power of CNN for classification task. Wang \textit{et al.} \cite{wang,wang2} proposed to encode times series as image using both the Gramian Angular Field and the Markov Transition Field. The obtained images are then concatenated and fed into a Tiled CNN network. The comparative study showed that such encoding improved the classification results of ECG signal. 
In \cite{emi}, similar encoding has been used for electromagnetic interference discharge classification. Local Binary Pattern (LBP) and the Local Phase Quantisation (LPQ) features are then extracted and classified using Random Forest method.
For supply-demand gap of mobile crowdsourcing, it is important to capture not only the intra-correlation, of the gap at a given location but also the inter-correlation. Indeed, the expected supply-demand is inherently influenced by the previous values in the closed past time frame. It is also affected by the supply-demand gap in older time frames. Hence, it is important to capture such information when forecasting the supply-demand gap. By transforming the time series into images, we will be able to capture the intrinsic properties of the gap time series to improve the forecasting performance. \newline
We propose to construct three types of images: Gramian Angular Summation Field (GASF), Gramian Angular Difference Field (GADF) \cite{ts_classification} and Recurrence Plot (REC) \cite{echman} from supply-demand gap time series. The intuition is to exploit the temporal correlation among gap data. GASF and GADF capture static information of time series by exploring the angular properties using trigonometry, while REC extracts the dynamic aspect of $Gap$ time series.\newline
The first step consists of segmenting $Gap$ into equal blocks of size $w$ and overlapping ratio $r$. Therefore, we are considering the changing properties of $Gap$ across time instead of one single observation. Then, each block, denoted $Gap_b$ is transformed to the polar coordinates after scaling in $[0,1]$. Each scaled ${gap_b}_i$, denoted $\widetilde{gap_b}_i$ is encoded to a pair of angle $\Psi_i$ and radisu $\rho_i$ by applying angular cosine function and considering the corresponding timestamp as the radius:
\begin{equation}
    \Psi_i = arccos(\widetilde{gap_b}_i);\quad \rho_i = \frac{i}{cst}
\end{equation}
$cst$ is a constant to control the polar plot span. Fig. \ref{polar} illustrates a polar plot of a gap time series.
Transforming $Gap_b$ into polar coordinates allows exploiting the angular property by using GASF and GADF. \newline
GASF is a matrix of the form:
\begin{equation}
    \begin{split}
        GASF_{i,j} & = cos(\Psi_i + \Psi_j)\\
        & =\widetilde{gap_b}^{Tr}  \widetilde{gap_b} - \Big( \sqrt{I-\widetilde{gap_b}^2}\Big)^{Tr}  \Big( \sqrt{I-\widetilde{gap_b}^2}\Big)
    \end{split}
\end{equation}
$GASF_{i,j;|i-j|=k}$ reflects the correlation by superposition of directions given time k. GADF is a matrix of the form:
\begin{equation}
    \begin{split}
        GADF_{i,j} & = sin(\Psi_i - \Psi_j)\\
        & = \Big(\sqrt{I - \widetilde{gap_b}^2}\Big)^{Tr} \widetilde{gap_b} - \widetilde{gap_b}^{Tr} \Big(\sqrt{I -\widetilde{gap_b}^2}\Big)    
    \end{split}
\end{equation}
$GADF_{i,j;|i-j|=k}$ reflects the correlation by difference of directions given time $k$. $I$ is the unit vector and $Tr$ is the transpose operator. We illustrate in Fig. \ref{gasf} the GASF and GADF  of two hours $Gap$ time series where each value is the overall supply-demand gap for each 10 min.\newline
In addition to the Gramian Angular Field, we encode the time series using the Recurrence Plot. Echman \textit{et al}. \cite{echman} investigated the recurrence of states in dynamic systems and proposed a tool that allows plotting the recurrence of these states in phase space. Usually the phase space is highly dimensional ($> 2$). Therefore, a projection into 2D or 3D is required to be able to visualize the recurrence states. Echman \textit{et al}. approach, known as Recurrence Plot, marks the state at time $i$ and at different time $j$ in two dimensional squared matrix with zero and one. Given a time series of length $w$, REC is mathematically expressed as:
\begin{equation}
    REC_{ij} = \mathcal{H}(\epsilon, ||{gap_b}_i - {gap_b}_j||) \quad i, j = 1 ... w
\end{equation}
$\mathcal{H}$ is the Heaviside function with parameter $\epsilon >0$:
\[
    \mathcal{H}(\epsilon,x)=\left\{
                \begin{array}{ll}
                  0 \quad \text{if}\; x\;<\; \epsilon\\
                  1 \quad \text{if}\; x\;>\; \epsilon \\
                \end{array}
              \right.
  \]
We illustrate the REC in Fig. \ref{rp}  with $\epsilon=0.5$.

\subsection{Deep residual network}
\begin{figure}[h!]
	\centering
	\includegraphics[scale=.6]{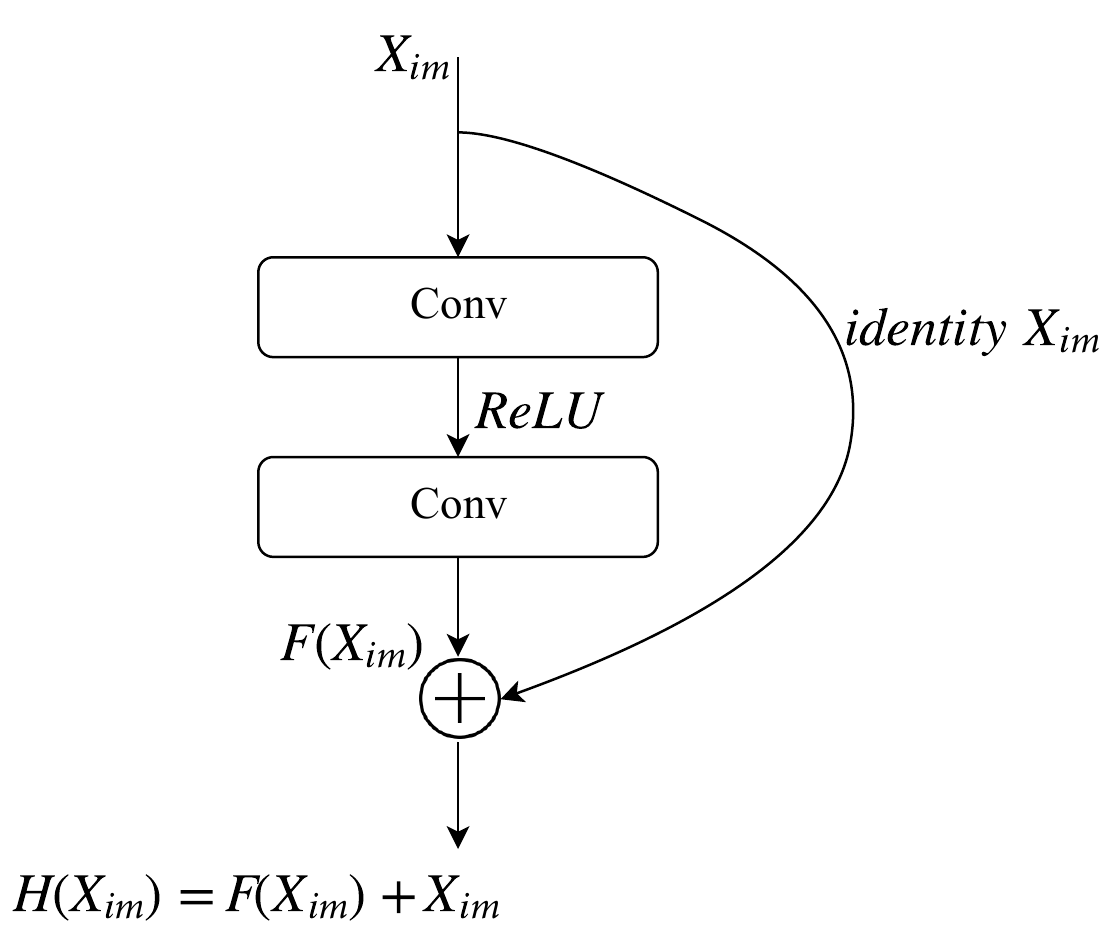}
	\caption{Residual block}
	\label{res_block}
\end{figure}

In this section, we give an overview of the residual network, a key component of Deep-Gap .\newline
In the following, $X_{im}$ refers to either $X_{GASF}$, $X_{GADF}$ or $X_{Rec}$. The pathway network is a deep CNN-based residual network \cite{resnet}. A residual block is illustrated in Fig. \ref{res_block}. It is a simple yet effective way that enables training very deep neural network while avoiding the notorious problem of vanishing/exploding gradient. Indeed, simply stacking network layers for the purpose of going deeper does not necessarily improve the results as the gradient would become very large or very low while optimizing the network parameters. This is obviously not an overfitting problem. Hence applying classic regularization techniques cannot improve results. To overcome this problem, authors in \cite{resnet},   proposed the residual learning concept. It is achieved through a residual block that maps the input $X_{im}$ to the output $F(X_{im})$ through two CNN layers. Then, the input is summed with the output through the identity connection. From training perspective, in worst case, the model would learn the identity of the input. A CNN layer applies a convolution operation on the input $a$ using a filter $\tau$ of size $n \times m$. At the $l^{th}$ layer and the $k^{th}$ neuron, the convolution operation is expressed as:
\begin{equation}
	 \sum_i^{n-1} \sum_j^{m-1} f( \tau_{i,j}a_{k-i,l-j})  +b 
	\label{cnn}
\end{equation} 
$b$ is a bias vector. $f$ is an activation function, typically a Rectified Linear Unit ($ReLU$), where:
\begin{equation}
    ReLU(x)=max(0,x)   
\end{equation}
The overall learning network of the pathway consists of CNN layer followed by $L$ residual blocks. The pathway extracts meaningful and discriminative features that can be used to forecast the $T+1$ gap.

\subsection{External network}\label{section_external_net}
Crowdsourced services supply and demand at a given area are realistically affected by multiple external factors. Indeed, the number of requests varies between weekends and weekdays. It is also affected by weather conditions as the gap is clearly different between snowy and clear conditions. In our experiments, we augment our learning framework with weather conditions and temperature data extracted from Accuweather service \footnote{\url{https://www.accuweather.com}} in addition to the day type (weekend, weekday). The first layer of the network is an Embedding layer. It is an approach used to represent text data using latent factors space. This representation allows circumventing the classic bag of words approach that leads to a high dimensional sparse vector. Instead, the Embedding layer enables representing words using dense vectors where a vector is the projection of a word into a continuous vector space. The Embedding layer is then followed by a fully connected layer.

\subsection{Network training}
The three pathways trained on $X_{GASF}$, $X_{GADF}$ and $X_{Rec}$ are followed by a concatenation layer whose output is $X_F= \{X_{GASF}, X_{GADF}, X_{Rec}\}$. The external layer output $X_{ext}$ is then also concatenated with $X_F$. The concatenation is fed into a fully connected layer whose output is the predicted supply-demand gap $X_{pred}$. The overall network is trained to minimize the loss between the predicted gap at $T+1$ and the actual value $X_{actual}$:
\begin{equation}
    \mathcal{L}(\theta) = ||X_{actual}-X_{pred}||^2
    \label{loss}
\end{equation}
In Eq. \ref{loss}, $\theta$ are the parameters of the overall network, to be learned. Algorithm \ref{PcP} details the training algorithm that optimizes the supply-demand gap forecasting network.
\begin{algorithm}[!h]
 \caption{Training algorithm}
 \label{PcP}
 \begin{algorithmic}[1]
\State \textbf{Input} $Gap$: time series of supply-demand gap; External data $D$, number of epochs, number of batches $n_b$
\State \textbf{Output:} Trained model with optimal set of  parameters $\theta$ 
\Statex \# \textit{Prepare data}
\State Create $GASF$, $GADF$ and $Rec.\; Plot$.
\State Create data instance $E =\{GASF, GADF, REC,D \}$
\Statex \# \textit{training the model}
\State \textbf{Repeat:}
\Statex \hspace{0.5cm} - Randomly select $n_b$ batch of $E$
\Statex \hspace{0.5cm} - Calculate $\theta$ that minimizes $L(\theta)$
\State \textbf{Until} stopping criterion is met
\end{algorithmic}
\end{algorithm}

\section{Experiments}
To validate the effectiveness of our approach, we conduct a set of experiments on different datasets and compare Deep-Gap against state-of-art algorithms.

\subsection{Datasets}
To the best of our knowledge, there are no crowdsourced service data that have been made public or reported in the literature. For this reason, we use similar data. In fact, recently, WiFi access has become available by several taxi services. The UberWiFi turns Uber's car into a mobile WiFi hotspot in Philadephia \footnote{\url{www.uber.com/blog/philadelphia/take-work-on-the-gowith-uberwifi.}}. Vinli offers WiFi on-board using 4G network for drivers and passengers \fnurl{}{https://www.uber.com/blog/new-orleans/stay-connected-on-your-ride}
\blfootnote{-with-free-wifi-powered-by-vinli/}. City Cabs \fnurl{}{www.citycabs.co.uk.} and CabWiFi \fnurl{}{www.cabwifi.com}, in London and Scotland, offer also free on-board WiFi access. Therefore, it is reasonable to use publicly available taxi data in this context.
We conduct our experiment on three taxi datasets:
\begin{itemize}
    \item \textbf{Yellow Taxi \fnurl{}{https://www1.nyc.gov/site/tlc/about/tlc-trip-record-data.page}:} These data contain records of New York yellow taxi in January 2018. The records include pick-up and drop-off time and pick-up and drop-off locations with their associated IDs. 
    \item \textbf{Porto Taxi \fnurl{}{https://www.kaggle.com/c/pkdd-15-predict-taxi-service-trajectory-i}:} consists of one year records of 442 taxis operating in Porto city, Portugal. The data provide trip records with time stamped geolocations and day type. We augment the data with weather data from Accuweather service.
    \item \textbf{Di-Tech Challenge \fnurl{}{https://ditech.didichuxing.com/en/}:} These data are provided by DiDi, the Chinese Taxi transportation company. The data contain the supply-demand gap of taxis at each district. We note that gap values are greater than zero. In addition, information about the weather are also provided. 
\end{itemize}

\subsection{Evaluation approach}
For both Yellow taxi and Porto taxi data, we consider each pick-up event as a demand record and each drop-off event as a supply, i.e. taxi has become available. Furthermore, to extract supply-demand record per region, we cluster locations into 17 clusters, representing the metropolitan municipalities. \newline
For fair comparison, we conduct two series of experiments. In the first one, prediction approaches are trained without external information (e.g. temperature, weather). We compare Deep-Gap against ARIMA (Autoregressive Integrated Moving Average), Gradient Boosting Decision Tree (GBDT), Long Short-Term Memory (LSTM)-based model, Bi-Directional LSTM (Bi-LSTM) \cite{bilstm} and Wavenet \cite{wavenet}. This setting is called baseline models. In the second experiment, we evaluate Deep-Gap against LSTM, Bi-LSTM and Wavenet. Each model is augmented with additional data as detailed in section \ref{section_external_net}. These are referred to as main models. We further analyze the effect of residual learning on Deep-Gap by training its baseline model in two versions:
with residual blocks and simple CNN layers instead of the residual blocks. In all experiments, we use 85\% of each data for training and 15\% for testing. We also set $L = 3$ for Porto Taxi, and Yellow Taxi, $L=2$ for DiDi Tech Challenge, $\epsilon = 0.5$, $w = 12$ samples, $r=1$ and $n=m= 5$. We report for each model the Root Mean Square Error (RMSE) on the test data:
\begin{equation}
    RMSE = \sqrt{\frac{1}{N} \sum\limits_{i=1}^N ( X_{actual}^i - X_{pred}^i )^2}
\end{equation}
\renewcommand*{\thefootnote}{\fnsymbol{footnote}}

\subsection{Forecasting performance}
\begin{figure}[h!]
	\centering
	\includegraphics[scale=.475]{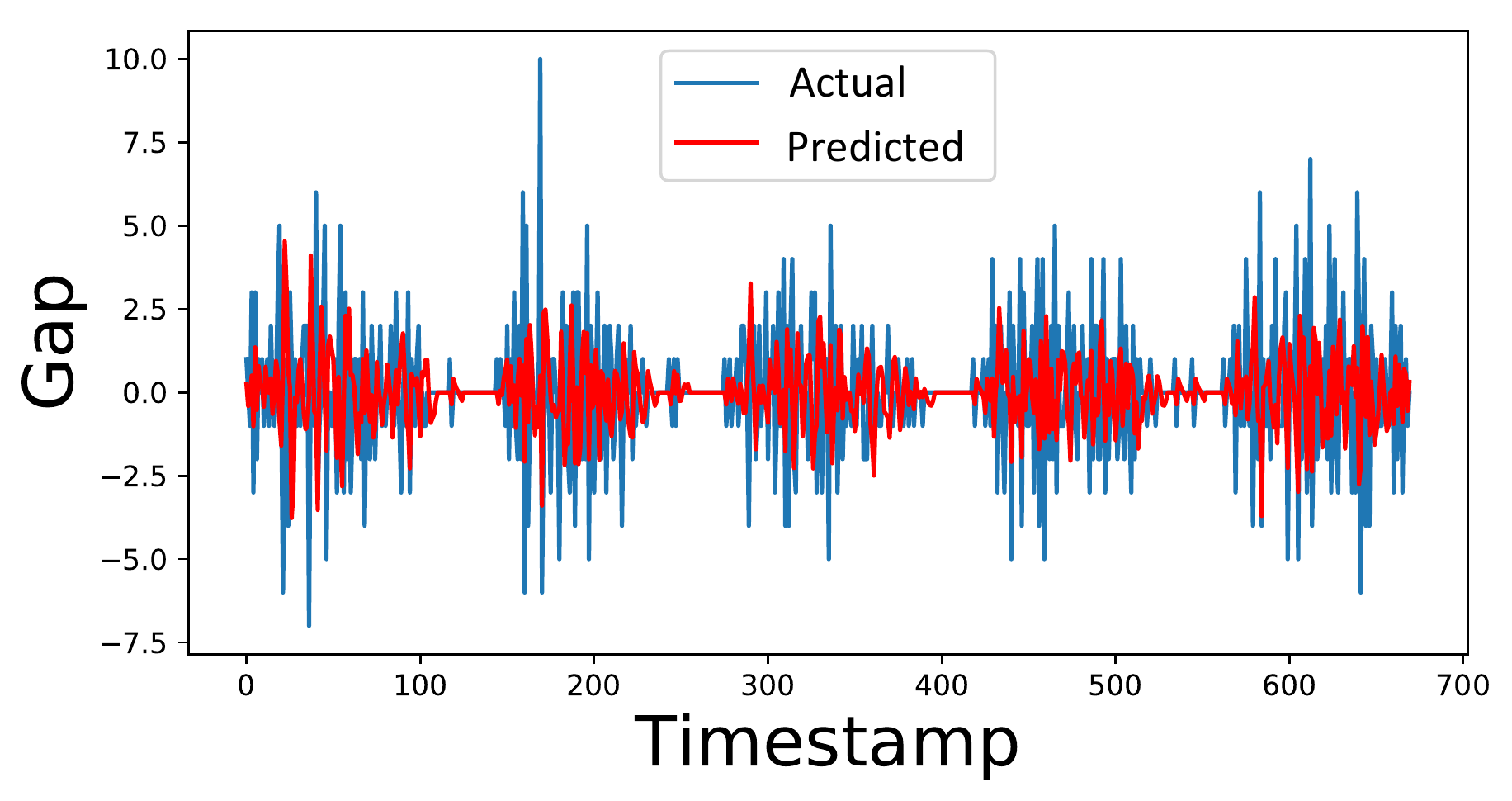}
	\caption{Yellow taxi supply-demand gap: Actual vs Prediction}
	\label{actual_vs_pred_yellow}
\end{figure}

\begin{figure}[h!]
	\centering
	\includegraphics[scale=.48]{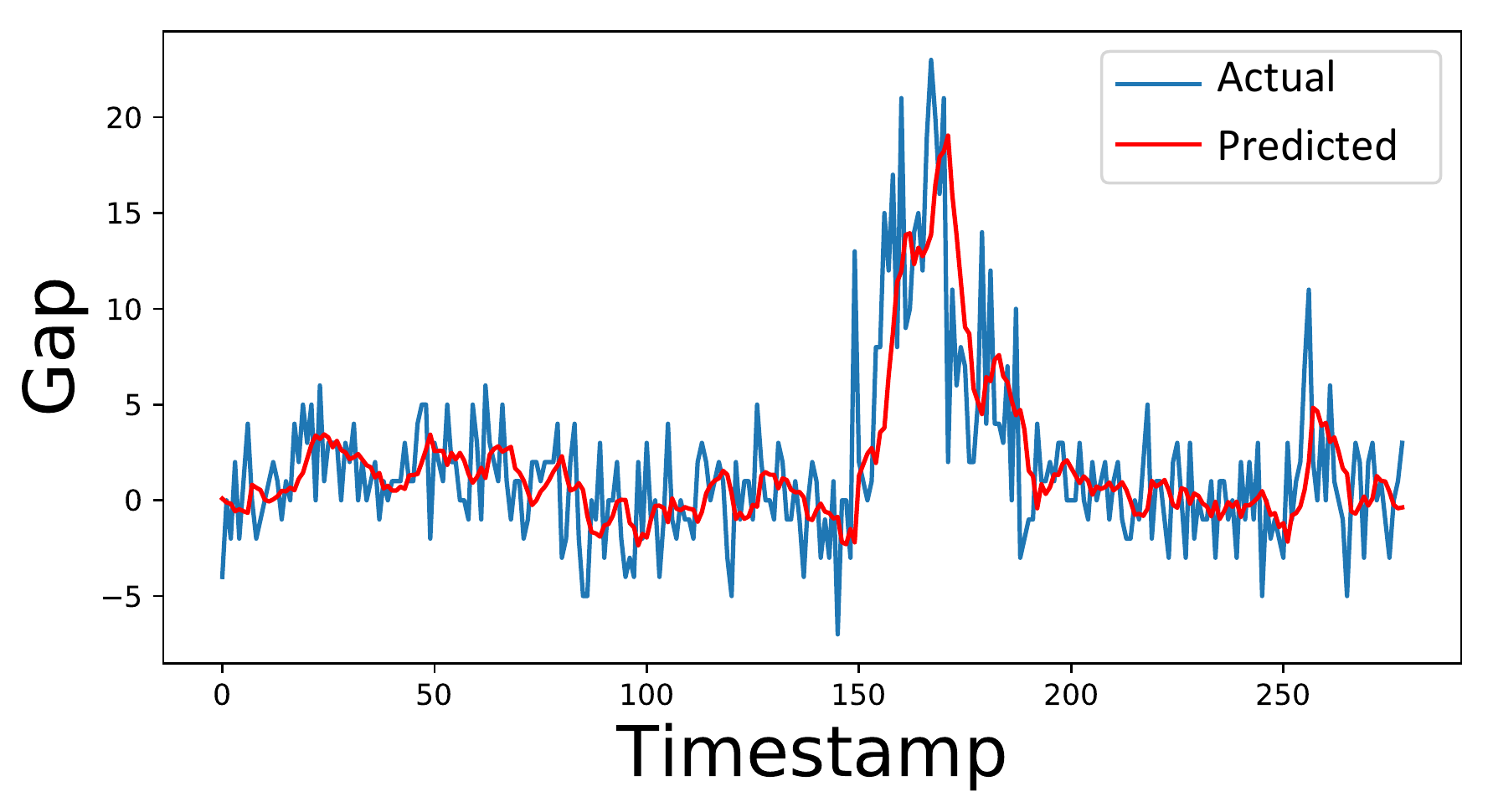}
	\caption{Porto taxi supply-demand gap: Actual vs Prediction}
	\label{actual_vs_pred_porto}
\end{figure}

\begin{figure}[h!]
	\centering
	\includegraphics[scale=.4]{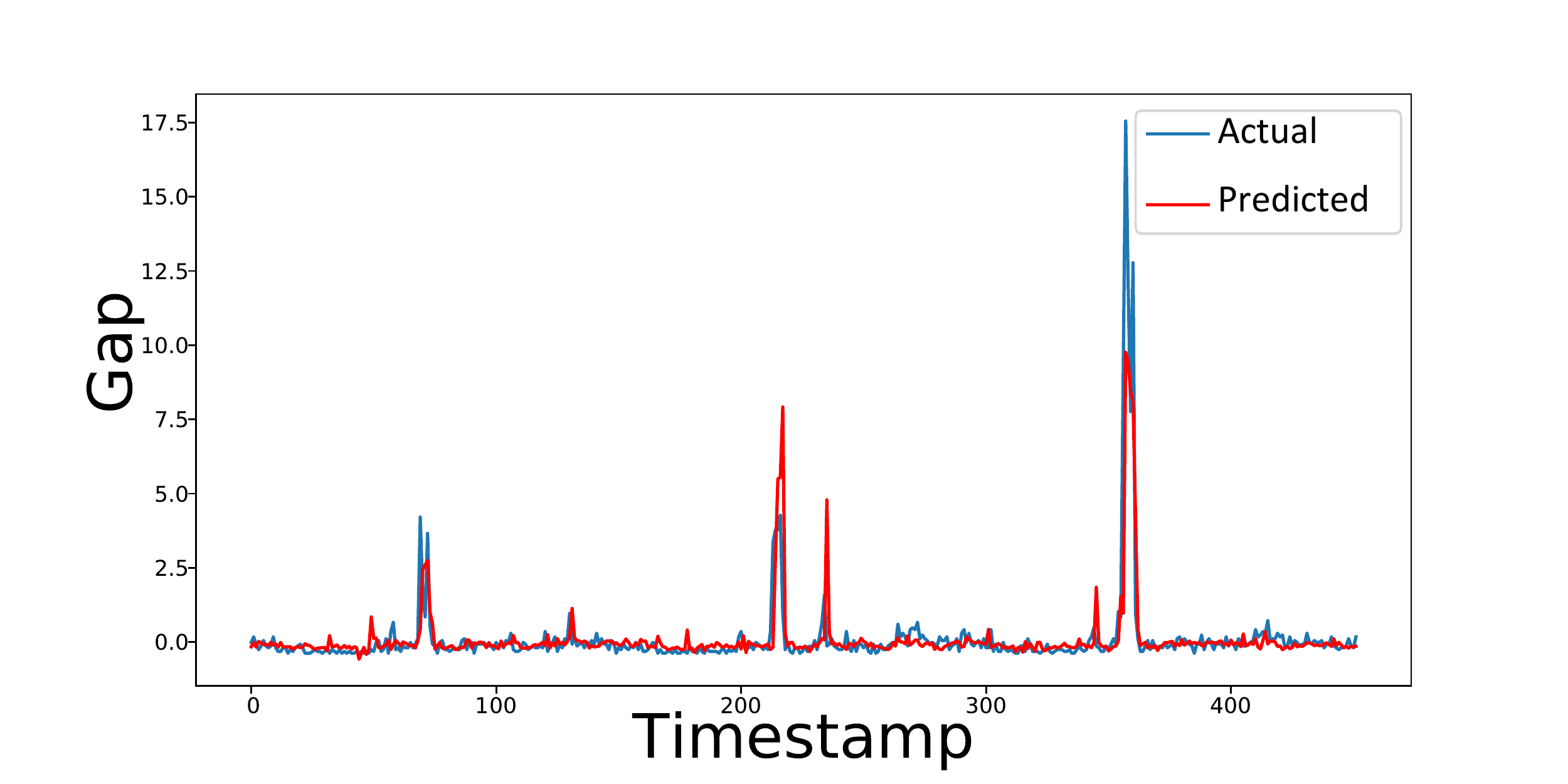}
	\caption{DiDi supply-demand gap: Actual vs Prediction}
	\label{actual_vs_pred_didi}
\end{figure}

Figures \ref{actual_vs_pred_yellow}, \ref{actual_vs_pred_porto} and \ref{actual_vs_pred_didi} show the actual supply-demand gap against the prediction of supply-demand datasets for district 1 from each data. The timestamp axis corresponds to the sampling time used to collect each data.
We notice that Deep-Gap with the external network is able to predict the gap variation particularly the sudden picks occurring across time. These particular sudden changes are very important to capture since service operators need to proactively anticipate  the pick in service demand which allows redirecting service providers to the district, thus maximizing the quality of experience and also profit.
Table  \ref{baseline_rmse} details the RMSE evaluation obtained by the forecasting approaches without external information, called baseline models. These findings show that Deep-Gap achieved the best performance with the least RMSE value while ARIMA achieved the least performance. Indeed, the improvements in terms of RMSEs are 22\%, 6\% and 11\% compared to the closest performance for Yellow Taxi, Porto Taxi and Didi Tech respectively. Table \ref{main_rmse} details RMSE performance of the main models. Results confirm that using additional information reduced the forecasting error of deep learning-based models. In addition,  Deep-Gap achieved the best performance with the lowest RMSE values for all evaluation datasets.
\begin{table}[h!]
\centering
\caption{Baseline models: RMSE  comparison }
\begin{tabular}{|c|c|c|c|}
\hline
\textbf{Approach} & Yellow Taxi & Porto Taxi & DiDi Tech  Challenge \\ \hline
ARIMA &5.71 &13.32 &15.21 \\ \hline
GBDT & 4.64&11.83 &16.13 \\ \hline
LSTM & 4.63 & 11.75 & 17.41 \\ \hline
Bi-LSTM & 4.60 & 11.74 & 15.07 \\ \hline
Wavenet & 4.62 & 11.73 & 15.58 \\ \hline
Deep-Gap & \textbf{3.58} & \textbf{10.97} & \textbf{13.37} \\ \hline
\end{tabular}
\label{baseline_rmse}
\end{table}

\begin{table}[h!]
\centering
\caption{Main models: RMSE comparison }
\begin{tabular}{|c|c|c|c|}
\hline
\textbf{Approach} & Yellow Taxi & Porto Taxi & DiDi Tech  Challenge\\ \hline
LSTM & 3.59 & 9.72 & 16.32 \\ \hline
Bi-LSTM & 3.58 & 9.69 & 14.57 \\ \hline
Wavenet & 3.61 & 9.72 & 15.21 \\ \hline
Deep-Gap & \textbf{2.88} & \textbf{7.95} & \textbf{11.15} \\ \hline
\end{tabular}
\label{main_rmse}
\end{table}
\begin{figure}[h!]
	\centering
	\includegraphics[scale=.25]{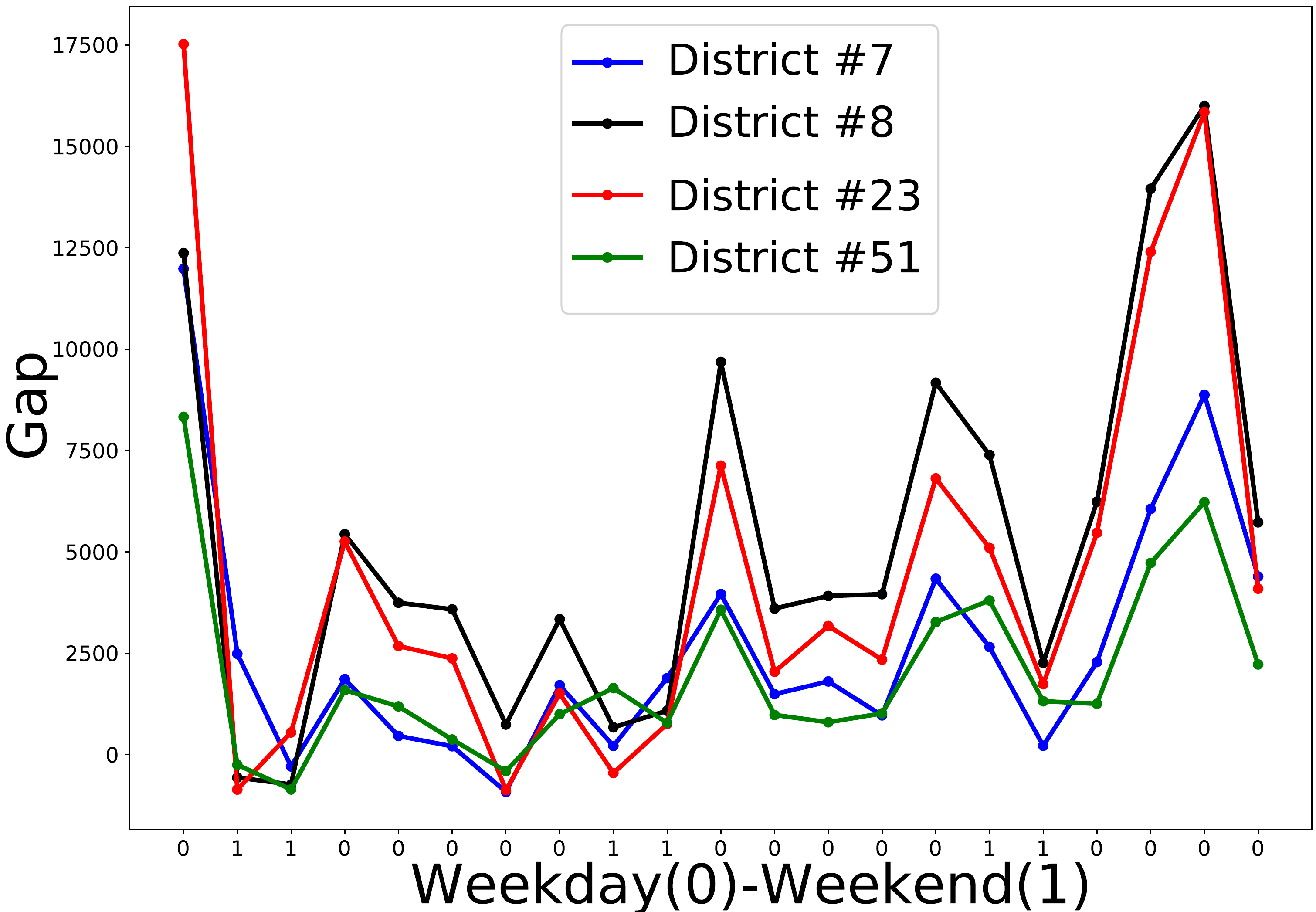}
	\caption{Supply-demand gap with respect to the day type}
	\label{actual_vs_pred}
\end{figure}
\noindent We illustrate in Fig. \ref{actual_vs_pred}, the variation of the DiDi supply-demand gap  with respect to the day type for different districts of Guangzhou, China. We can clearly notice that the gap range of values is different between the day types. Indeed, the gap is very high during weekdays compared to weekends. Thus, it is important to capture such information in any gap prediction model.

\subsection{Effect of Residual block}
\begin{figure}[h!]
	\centering
	\includegraphics[scale=.24]{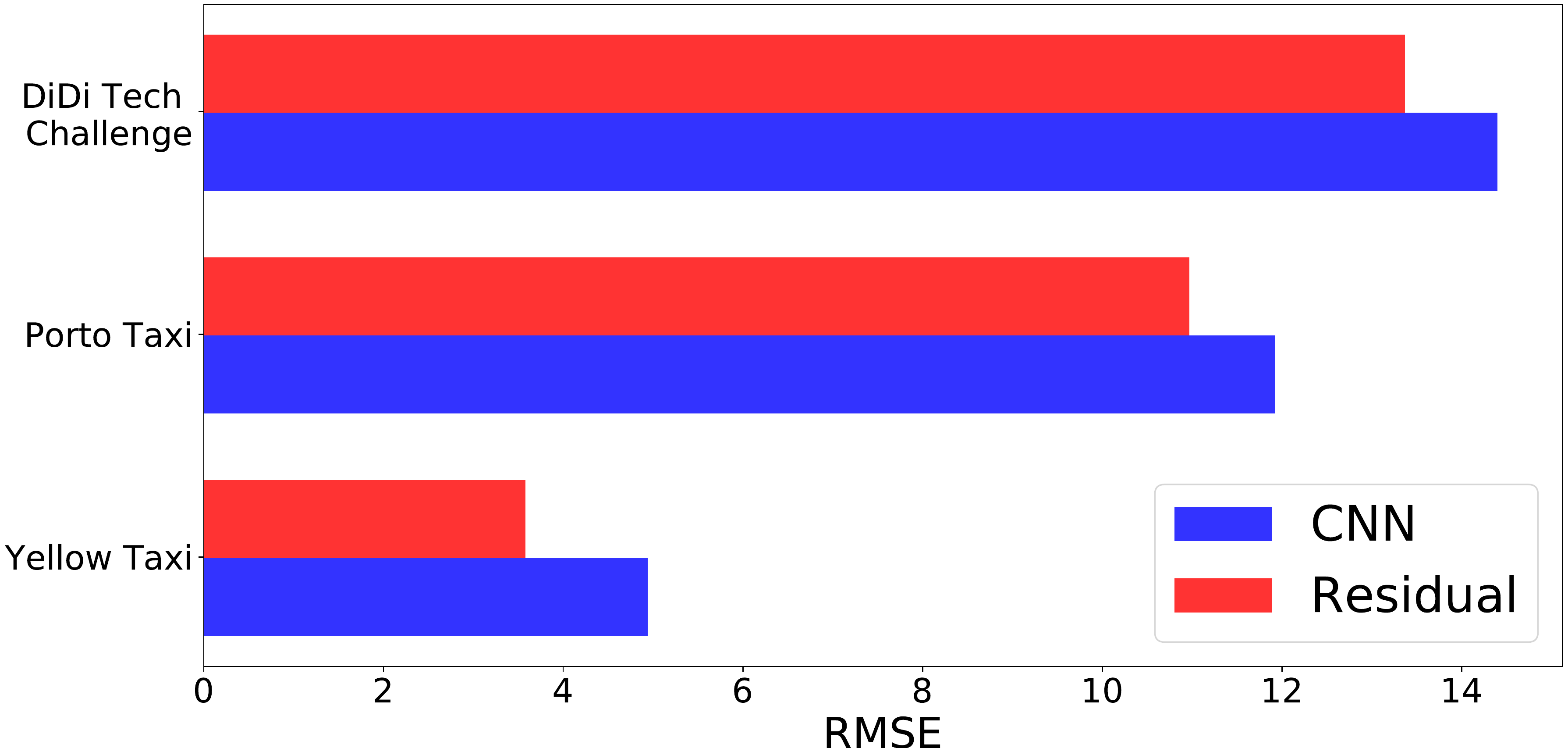}
	\caption{Residual vs CNN: RMSE performance}
	\label{cnn_vs_residual}
\end{figure}

We analyze in this section the effect of adopting residual learning. We evaluate the performance of Deep-Gap by comparing the deep neural network with Deep-Gap where we replace residual blocks with CNN layers without skip connections. The performance comparison is depicted in Fig. \ref{cnn_vs_residual}. The results show that residual-based model achieved the best performance. This confirms that residual blocks provide better learning capacity.
\subsection{Discussion}
Deep-Gap showed competitive performance for supply-demand gap forecasting. Visual inspection of the predicted values compared to the actual gap values shows that Deep-gap tends to follow the trend of the actual supply-demand gap and successfully captures the sudden peaks. We notice also that Deep-Gap tends to smooth the prediction values. Quantitatively, the proposed learning approach outperforms well known forecasting methods in terms of prediction accuracy. Deep-Gap can easily accommodate external data that certainly influence the variation of supply-demand gap such as the weather conditions. It is worth noting that it is important to train Deep-Gap on sufficiently large data as any deep learning task. Furthermore, it is important to collect mobile crowdsourcing data to provide better understanding of the crowd commitment to the crowdsourcing task.
\section{Conclusion}
We proposed Deep-Gap, a deep learning time series forecasting model to predict the supply-demand gap of crowdsourced services. The time series of service supply-demand  are encoded as images and fed into a series of residual blocks of CNNs. The proposed network is augmented with additional external data, such as the weather conditions, that realistically impact the supply-demand gap at a given region. The comparison with other forecasting approaches confirmed that Deep-Gap provides an improved prediction performance. In future work, we will investigate neural architecture search techniques to determine the optimal learning network for times series images.

\section*{Acknowledgment}
This publication was made possible by NPRP grant \# NPRP9-224-1-049 from the Qatar National Research Fund (a member of Qatar Foundation). The statements made herein are solely the responsibility of the authors.

\end{document}